\documentclass[aps,prd,reprint,nofootinbib,superscriptaddress]{revtex4-1}
\pdfoutput=1

\usepackage{setspace}
\usepackage{booktabs}
\usepackage{graphicx}
\usepackage{amsmath}
\usepackage{amsfonts}
\usepackage{amssymb}
\usepackage{rotating}
\usepackage[dvipsnames]{xcolor} 
\usepackage{tikz}
\usepackage{float}
\usepackage{multirow}
\usepackage{slashed}
\usepackage{hyperref} 
\usepackage{changes}

\hypersetup{colorlinks=true, linkcolor=OrangeRed, urlcolor=OrangeRed, citecolor=OrangeRed}
 
\newcommand{\AddrMPI}{Max-Planck-Institut f\"ur Kernphysik,\\ Saupfercheckweg 1, 69117 Heidelberg, Germany}

\newcommand{\AddrFCFM}{CIFFU and Facultad de Ciencias F\'isico-Matem\'aticas, \\ Benem\'erita
Universidad Aut\'onoma de Puebla,  Puebla,  Pue.~72000,  M\'exico}

\begin{document}

\title{{\color{OrangeRed}Natural 2HDMs without FCNCs}}

\author{J. L. Diaz-Cruz}
\email{jldiaz@fcfm.buap.mx}
\affiliation{\AddrFCFM}

\author{U. J. Saldana-Salazar}
\email{saldana@mpi-hd.mpg.de}
\affiliation{\AddrMPI}

\author{K. M. Tame-Narvaez}
\email{tame@mpi-hd.mpg.de}
\affiliation{\AddrMPI}

\author{V. T. Tenorth}
\email{tenorth@mpi-hd.mpg.de}
\affiliation{\AddrMPI}

\begin{abstract}
\noindent
Motivated by the fermion mass hierarchy we study the phenomenology of two flavorful two-Higgs-doublet model (2HDM) scenarios. By virtue of the flavor or singular alignment ansatz it is possible to link the mass of a subset of fermions to the vacuum-expectation-value (VEV) of a unique Higgs doublet and to simultaneously avoid flavor-changing-neutral-currents at tree-level.
We explicitly construct two models called Type-A and B. There, either the top quark alone or all third generation fermions couple to the doublet with the larger VEV. The other fermions acquire their masses through the small VEV of the other doublet.
Thus, more natural values for the Yukawa couplings can be obtained.
The main differences between these models and conventional ones are studied including a discussion of both their structure and phenomenological consequences.
In particular, as distinctive deviations for the Yukawa couplings of the light fermions are predicted we discuss possible tests at the LHC based on searches for $h\to J/\Psi + \gamma$, $h\to\mu\mu$, and heavy scalar resonances decaying to muon pairs. 
We find that for a wide region of parameter space this specific set of signatures can be used to distinguish among the new proposed types and the conventional ones.
\end{abstract}

\maketitle

\section{INTRODUCTION}
\label{sec:intro}
\noindent
Flavor-changing-neutral-current processes (FCNC) have been experimentally observed to be strongly suppressed. Their smallness in the Standard Model (SM) has long been understood through the Glashow--Iliopoulos--Maiani mechanism~\cite{Glashow:1970gm}. However, its simplicity requires considering only one Higgs doublet; once two scalar doublets are assumed potential tree-level flavor transitions are expected. Those FCNCs are mediated by the linear combination of the neutral components of the doublets. Similarly, either assuming the SM to be an effective field theory or including only a gauge singlet scalar, the addition of higher dimensional operators will produce undesirable flavor violating effects. To overcome this situation one must invoke further assumptions. This creates many alternative paths, and gives place to some ambiguity. Generally speaking, going beyond the standard scenario always new FCNC sources arise. In particular, for any multi-Higgs extension those arise already at tree-level. We may then ask: why should we include more Higgs fields if the single-Higgs picture is already naturally consistent with small FCNCs? 

To answer the previous question we need to consider which other fundamental aspects of the SM can be addressed by extending the number of Higgs doublets, and investigate what would be the main phenomenological consequences of such scheme. In this respect, through naturalness arguments~\cite{Dirac:1938mt,tHooft:1979rat} fermion masses give rise to one of the still unsolved mysteries in the SM: the problem of mass hierarchy~\cite{Weinberg:1977hb}. An application of Dirac's naturalness criteria~\cite{Dirac:1938mt} requires the observation of all fermion masses around the electroweak (EW) scale, $m_f = y_f v$ with $y_f \sim {\cal O}(1)$ and $v = 174 \text{ GeV}$. This is because the aforementioned criteria requires all dimensionless couplings of a theory to be of order one to be considered natural.
Curiously, it is only the top quark which satisfies it in the SM.
However, the \textit{unnaturalness} of the lighter fermions, which can be described as: 
\begin{equation}
    y_t \sim 1 \gg y_f \qquad (f = c,b,s,u,d,\tau,\mu,e)\;,
\end{equation} 
could still be called natural according to 't Hooft's criteria~\cite{tHooft:1979rat}. Namely, if the system acquires a symmetry by setting the tiny Yukawa couplings to zero, it can be called natural.
This is already the case in the SM. However, a caveat exist: there is no common Yukawa parameter among this set of masses that could simultaneously bring all of them to zero. However, in a multi-scalar theory with at least two vacuum-expectation-values (VEVs), $v_1$ and $v_2$, one could instead use the VEVs to accomplish it, namely
\begin{equation}
    m_t = y_t v_2 \qquad \text{and} \qquad m_f = y_f v_1\;,
\end{equation}
where $f=c,b,s,u,d,\tau,\mu,e$ and $v\sim v_2 \gg v_1$.
Here, besides $y_t$ being of order one as in the SM other Yukawa couplings could also be of the same order, e.g.\,for those fermions with masses around $v_1\sim{\cal O}(1\,\text{GeV})$. 
Nonetheless, this is not a general feature of multi-scalar theories, as fermion masses normally feature a linear combination of VEVs, e.g.\,for two scalar doublets $m_f = y_{f,1} v_1 + y_{f,2} v_2$.
Therefore, our main motivation in this work is to build models where through a common VEV sets of light masses could be linked.
This is certainly an ambitious goal. First, we need to check if such models satisfy all current theoretical and phenomenological constraints.
Moreover, as previously mentioned, this small VEV should be consistent with 't Hooft's criteria of naturalness. This means that its non-zero value should come from the breaking of a symmetry. Furthermore, we note that due to $v_1^2$ being sufficiently smaller than $v_2^2$ such that $v_2 \sim v$, even in the case where no additional Yukawa coupling are of order one the scalar sector would still fulfill 't Hooft's criteria for naturalness.
Hence, in the two aforementioned meanings the following theories are called natural.

The simplest renormalizable choice along this line is to extend the scalar sector to contain two Higgs doublets ($\Phi_1$, $\Phi_2$), the so-called two-Higgs doublet models (2HDMs)~\cite{Branco:2011iw}. 
Often, the Yukawa sector of these models includes a family or flavor universal $\mathbb{Z}_2$-parity assignments which also operates on the Higgs potential.
This is understandable as natural flavor conserving (NFC) theories demand it~\cite{Paschos:1976ay,Glashow:1976nt}. Nonetheless, more recently some family non-universal setups have been investigated~\cite{Ibarra:2014fla,Bauer:2015kzy,Ghosh:2015gpa, Altmannshofer:2015esa,Botella:2016krk,Altmannshofer:2017uvs,Altmannshofer:2018bch,Chulia:2019evx,Nomura:2019dhw}. In general, all of these setups contain tree-level FCNCs.
As mentioned before in this case one must invoke different mechanisms to ensure the suppression of such currents.
The models proposed here assume their parameters to be in a certain region of flavor space. There, all Yukawa matrices become diagonal in the mass basis irrespective of the fact that they were initially not proportional to the mass matrix. 
This ansatz is known as Yukawa alignment~\cite{Pich:2009sp} in its flavor universal form, or general flavor (also called singular) alignment in its flavor non-universal one~\cite{Penuelas:2017ikk,Rodejohann:2019izm}. The two models to be proposed here are examples of the latter ansatz. 
\\

The outline of this paper is as follows. In Sec.~\ref{sec:2hdms} we review the main theoretical constraints on 2HDMs with a softly-broken $\mathbb{Z}_2$ symmetry, and show how to build a hierarchical VEV alignment. Afterwards, in Sec.~\ref{sec:SA} we briefly revisit the singular alignment ansatz which guarantees the absence of tree-level FCNCs. We then apply it in Sec.~\ref{sec:models} in a judicious manner to construct two new types of 2HDMs whose scalar mediated fermion interactions conserve flavor at tree-level. Thereafter, in Sec.~\ref{sec:pheno} we define benchmark scenarios that will simplify our analysis and the main phenomenological constraints which are relevant to them. In Sec.~\ref{sec:results} we discuss the most critical differences between those new models and conventional 2HDMs, and explore phenomenological consequences that can test our models. 
We finally conclude in Sec.~\ref{sec:conc}. 

\section{THE 2HDM, HIERARCHICAL VEVs AND THEORETICAL CONSTRAINTS}
\label{sec:2hdms}\noindent
The components of the two Higgs doublets  are written as  follows:
\begin{equation}
    \Phi_j = \begin{pmatrix}
    \phi_j^+ \\
    v_j + \phi_j^0   \end{pmatrix} \qquad
    (j = 1,2) \;,
    \label{eq:scalar}
\end{equation}
where $v_j$ represents the VEV that could in general be complex, e.g.~in a Charge-Parity (CP) violating potential. Introducing a $\mathbb{Z}_2$ symmetry reduces the arbitrariness in the Higgs potential. In addition, if wisely applied to the Yukawa sector, it guarantees the absence of tree-level FCNCs~\cite{Paschos:1976ay,Glashow:1976nt} (later discussed in more detail). We conventionally assign the $\mathbb{Z}_2$-parities
\begin{equation}
    \Phi_2 \to + \Phi_2 \qquad \text{and} \qquad
    \Phi_1 \to - \Phi_1 \;.
\end{equation}
Then the most general $\mathbb{Z}_2$-invariant scalar potential is given by
\begin{align} \label{eq:2HDMZ2}
    V_\text{2HDM}^{\mathbb{Z}_2}  =& \sum_{x=1,2} \left[ m^2_{xx} (\Phi^\dagger_x \Phi_x) +\frac{\lambda_x}{2} (\Phi^\dagger_x \Phi_x)^2 \right] \nonumber\\
    +& \lambda_3 (\Phi^\dagger_1 \Phi_1) (\Phi^\dagger_2 \Phi_2) +\lambda_4 (\Phi^\dagger_1 \Phi_2) (\Phi^\dagger_2 \Phi_1) \\
    +& \frac{1}{2} \left[ \lambda_5(\Phi^\dagger_1 \Phi_2)^2 + \lambda_5^*(\Phi^\dagger_2 \Phi_1)^2 \right] \;.\nonumber
\end{align}
Notice that by demanding hermiticity, the potential keeps $\lambda_5$ as the only complex coefficient while $m_{11}^2$, $m_{22}^2$, and $\lambda_{1,2,3,4}$ are real. By a phase redefinition the complex phase of $\lambda_5$ can be turned to zero without loss of generality. Therefore, only seven parameters are physical, and the potential is CP-symmetric. For a thorough assessment of two-Higgs-doublet models (2HDMs) please refer to~\cite{Branco:2011iw} and for more recent reviews to~\cite{Bhattacharyya:2015nca,Diaz-Cruz:2019vka}. 

In order to create a hierarchy among the VEVs, we guarantee that in a first stage only $\Phi_2$ develops a VEV by assuming
\begin{equation}
    m_{22}^2 <0 \qquad \text{and} \qquad m_{11}^2>0 \;.
\end{equation}
Therefore, the $\mathbb{Z}_2$ symmetry is preserved, and  
\begin{equation} \label{eq:VEV2}
    v_2 = \sqrt{\frac{-m_{22}^2}{\lambda_2}} \;,
\end{equation}
while $v_1 = 0$.
The second stage requires to softly-break the symmetry by adding the terms
\begin{equation}
    - m_{12}^2 (\Phi_2^\dagger \Phi_1 + \Phi_1^\dagger \Phi_2)
\end{equation}
to Eq.~\eqref{eq:2HDMZ2}. Choosing $m_{12}^2$ to be real ensures that these terms preserve CP.
If the condition $m_{12}^2 t_\beta \gg \lambda_1 v_1^2$ with $t_\beta = v_2 / v_1$ is met $\Phi_2$ induces a small VEV to $\Phi_1$ of the form
\begin{equation} \label{eq:VEV1}
    v_1 \simeq \frac{m_{12}^2 v_2 }{m_{11}^2 + \lambda_{345}v_2^2} \;,
\end{equation}
where $\lambda_{345} \equiv \lambda_3 + \lambda_4 + \lambda_5$. One can show that in this case the heavy scalar masses are above the EW scale.
For completeness, we add here the two minimization conditions:
\begin{equation}
    \begin{gathered}
   m_{22}^2 v_2 = m_{12}^2 v_1 -\lambda_2 v_2^3 -\lambda_{345} v_2 v_1^2 \;,\\
   m_{11}^2 v_1 = m_{12}^2 v_2 -\lambda_1 v_1^3 -\lambda_{345} v_1 v_2^2 \;,
    \end{gathered}
\end{equation} 
from which Eqs.~\eqref{eq:VEV2} and~\eqref{eq:VEV1} are derived.

For the sake  of illustration, the small VEV can be estimated by assuming $\lambda_{345} \sim {\cal O}(1)$ and $m_{11} \sim v_2$, thus obtaining
\begin{equation}
    v_1 \sim \frac{m_{12}^2}{v} \;.
\end{equation}
Hence, if $m_{12} \sim {\cal O}(10\,\text{GeV})$ then $v_1 \sim {\cal O}(1\,\text{GeV})$.
The smallness of $v_1$ is natural as setting it to zero one recovers the initial $\mathbb{Z}_2$ symmetry.
Now, as both VEVs contribute to the $W$-boson mass they satisfy
\begin{equation}
    v^2 = v_1^2 + v_2^2 = (174\,\text{GeV})^2 \;.
\end{equation}
It is straightforward to realize that we still expect the large VEV to be close to the EW scale, i.e. $v_2 \approx v$.

The physical states of the CP-symmetric potential are two CP-even ($h,\,H$), one CP-odd ($A$) neutral scalars, and a pair of charged scalars ($H^\pm$). The transition from the interaction 
($\text{Re}(\phi_{1,2}^0),\, \text{Im}(\phi_{1,2}^0),\, \phi^\pm_{1,2}$)
to the mass basis ($h,H,A,H^\pm,G^0,G^\pm$) depends on two mixing angles ($\alpha,\beta$),
\begin{align}
\begin{aligned}
    \begin{pmatrix}
    h \\ H
    \end{pmatrix}
    & = \begin{pmatrix}
    \cos\alpha & -\sin\alpha \\
    \sin\alpha & \cos\alpha
    \end{pmatrix}
    \begin{pmatrix}
    \text{Re}(\phi_2^0) \\  \text{Re}(\phi_1^0)
    \end{pmatrix}\;,
    \\
    \begin{pmatrix}
    A \\ G^0
    \end{pmatrix}
    & = \begin{pmatrix}
    \cos\beta & -\sin\beta \\
    \sin\beta & \cos\beta
    \end{pmatrix}
    \begin{pmatrix}
    \text{Im}(\phi_2^0) \\  \text{Im}(\phi_1^0)
    \end{pmatrix} \;,
    \\
    \begin{pmatrix}
    H^+ \\ G^+
    \end{pmatrix}
    & = \begin{pmatrix}
    \cos\beta & -\sin\beta \\
    \sin\beta & \cos\beta
    \end{pmatrix}
    \begin{pmatrix}
    \phi^+_2 \\ \phi_1^+
    \end{pmatrix} \;,
\end{aligned}
    \label{eq:rotation}
\end{align}
where $G^0$ and $G^+$ denote the required two massless SM Goldstone bosons.
In the following, we denote the SM-like Higgs as $h$ with the mass $m_h = 125 \text{ GeV}$.

Through the invariants of the scalar mass matrices and
\begin{equation}
    t_{2\alpha} = \frac{2(v^2 \lambda_{345}\,s_{2\beta} -m_{12}^2)}{m_{12}^2 (t_\beta - t^{-1}_\beta)+2v^2 (c_\beta^2 \lambda_1 - s_\beta^2 \lambda_2)} \; ,
\end{equation}
the quartic couplings of the scalar potential can be expressed in terms of the Higgs mass eigenvalues ~\cite{Arnan:2017lxi,Arcadi:2019lka,Arco:2020ucn} 
\begin{align} \label{eq:lambdasmasses}
\begin{aligned}
    \lambda_1 &= \frac{1}{2v^2 c_\beta^2}\Big(m_h^2 c_\alpha^2 + m_H^2 s_\alpha^2 - M^2 s^2_\beta \Big) \;, \\
    \lambda_2 &= \frac{1}{2v^2 s_\beta^2}\Big(m_h^2 s_\alpha^2 + m_H^2 c_\alpha^2 - M^2 c^2_\beta\Big) \;, \\
    \lambda_3 &= \frac{1}{2v^2}\Big[\frac{s_{2\alpha}}{s_{2\beta}}(m_H^2-m_h^2) + 2 m_{H^\pm} - M^2\Big] \;, \\
    \lambda_4 &= \frac{1}{2v^2}\Big(M^2 + m_A^2 - 2 m_{H^\pm}^2\Big) \;, \\ 
    \lambda_5 &= \frac{1}{2v^2}\Big(M^2 - m_A^2\Big) \;,
\end{aligned}
\end{align}
where $M^2 \equiv 2 m^2_{12}/s_{2\beta}$. We note that $\lambda_{345} = [M^2 + (m_H^2-m_h^2)s_{2\alpha}/s_{2\beta}]/(2v^2)$. Moreover, for the scalar potential to be bounded from below (BFB) the quartic couplings should fulfill \cite{Klimenko:1984qx,Gunion:2002zf}
\begin{equation} \label{eq:stabilityrequ}
\begin{gathered}
    \lambda_{1,2} \geq 0 \;, \qquad
    \lambda_3 \geq -\sqrt{\lambda_1 \lambda_2}\;, \\
    \lambda_3 + \lambda_4 - |\lambda_5| \geq - \sqrt{\lambda_1 \lambda_2} \;.
\end{gathered}
\end{equation}
From requiring unitarity and perturbativity the coefficients have to satisfy the following relations~\cite{Akeroyd:2000wc,Bhattacharyya:2015nca}
\begin{align} \label{eq:2HDMS_unitarity}
\begin{split}
     |\lambda_3 + 2 \lambda_4 \pm 3 \lambda_5| \leq 16 \pi \;,\\
     |\lambda_3 \pm \lambda_4| \leq 16 \pi  \;, \qquad |\lambda_3 \pm \lambda_5| \leq 16 \pi \,,\\
     \Big|\frac{1}{2}\left( \lambda_1+\lambda_2\pm\sqrt{(\lambda_1 - \lambda_2 )^2+4 \lambda_4^2}\right)\Big| \leq 16 \pi \;, \\
     \Big|\frac{1}{2}\left( \lambda_1+\lambda_2\pm\sqrt{(\lambda_1 - \lambda_2 )^2+4 \lambda_5^2}\right)\Big| \leq 16 \pi \;, \\
     \Big|\frac{1}{2}\left( 3\lambda_1+3\lambda_2\pm\sqrt{9(\lambda_1 - \lambda_2 )^2+4 (2\lambda_3+\lambda_4)^2}\right)\Big|\leq 16 \pi \;.
\end{split}
\end{align}
These constraints indirectly ensure that the potential remains perturbative up to very high scales. Any additional constraint on the sizes of the $\lambda_i$ will make the analysis more restrictive. 

For last, the requirement that should be met in order to guarantee that the minimum is a global one is~\cite{Barroso:2013awa}
\begin{equation}
    m_{12}^2 \left(m_{11}^2 - m_{22}^2 \sqrt{\frac{\lambda_1}{\lambda_2}}\right)\left(t_\beta - \sqrt[4]{\frac{\lambda_1}{\lambda_2}} \right) > 0\;.
\end{equation}

\section{SINGULAR ALIGNMENT OR HOW TO AVOID TREE-LEVEL FCNCS}
\label{sec:SA}
\noindent
In general, Yukawa interactions in 2HDMs without a $\mathbb{Z}_2$ symmetry imply two independent contributions to the fermion mass matrices
\begin{equation} \label{eq:MassMatrix2HDM}
    {\bf M} = v_1 {\bf Y}_1 + v_2 {\bf Y}_2 \;.
\end{equation}
As the diagonalization of the mass matrices do not imply simultaneous diagonalization of both Yukawa matrices tree-level FCNCs occur. This scenario is typically called Type-III. However, once we introduce a parity symmetry in the scalar sector two scenarios arise for the mass matrices: $i)$ only one Yukawa matrix contributes, i.e. $\mathbb{Z}_2$ is flavor universal, or $ii)$ certain columns (or rows, depending on the parity assignment) of both Yukawa matrices contribute, i.e. $\mathbb{Z}_2$ is flavor non-universal. The former case allows for four different combinations to assign the charged fermions to $\Phi_{1,2}$.
Those are conventionally called Type-I, II, X, Y, and are shown in Table~\ref{tab:2HDM types}. The four combinations prohibit tree-level FCNCs and are called NFC theories~\cite{Paschos:1976ay,Glashow:1976nt}.

On an equal footing, two scenarios without tree-level FCNCs are possible: i) the two Yukawa matrices are proportional to each other, or ii) the rank one matrices corresponding to the eigenvalues of each Yukawa matrix are proportional. The former ansatz is called Yukawa alignment~\cite{Pich:2009sp} whereas the latter one is its generalized version~\cite{Penuelas:2017ikk} also known as singular alignment~\cite{Rodejohann:2019izm}.

When considering NFC theories the $\mathbb{Z}_2$ symmetry is usually applied to the right-handed (RH) fermion fields but it could also be applied to the left-handed ones. To illustrate the previous discussion we consider the $\mathbb{Z}_2$ assignments for the Type-II scenario,
\begin{equation}
\begin{gathered}
    d_{i,R} \to - \,d_{i,R} \;, \qquad e_{i,R} \to - \,e_{i,R} \;, \\
    u_{i,R} \to + \,u_{i,R} \;,
\end{gathered}
\end{equation}
with $i=1,2,3$. All left-handed fermions are even under the parity symmetry. Then, $\Phi_2$ can only couple to up-type quarks, while $\Phi_1$ couples to down-type quarks and charged leptons, shown in the second column of Table~\ref{tab:2HDM types}.
Interestingly, the four types can be encompassed via the Yukawa-alignment~\cite{Pich:2009sp} ansatz 
\begin{equation} \label{eq:YukAlig}
    {\bf Y}_1 \propto {\bf Y}_2 \;,
\end{equation}
as well as through its generalization in~\cite{Penuelas:2017ikk,Rodejohann:2019izm}.

In the latter ansatz~\cite{Penuelas:2017ikk} the $Z_2$ symmetry is applied family non-universally as each Higgs doublet couples only to a given generation of each fermion family.
In one literature example~\cite{Botella:2016krk} $\Phi_2$ couples only to the third generation while $\Phi_1$ to the first and second ones.
Compared to the first case (global application of the symmetry) a general feature of these scenarios is that they give rise to tree-level FCNCs. Therefore, one must find ways to sufficiently suppress them. 

\begin{table}[t]    \centering 
    \begin{tabular}{lcccc}
        Type- & I & II & X & Y \\ \hline\hline
        $u_{i}$ & $\Phi_2$ & $\Phi_2$ & $\Phi_2$ & $\Phi_2$ \\
        $d_{i}$ & $\Phi_2$ & $\Phi_1$ & $\Phi_2$ & $\Phi_1$ \\
        $\ell_{i}$ & $\Phi_2$ & $\Phi_1$ & $\Phi_1$ & $\Phi_2$
    \end{tabular}
    \caption{The four different types of 2HDMs with NFC. The allowed couplings between each fermion and a certain Higgs doublet are imposed by a group symmetry, e.g. a $\mathbb{Z}_2$. 
    Here, either the right-handed or left-handed components obtain a non-trivial charge assignment.}
    \label{tab:2HDM types}
\end{table}

The singular alignment ansatz introduced in~\cite{Rodejohann:2019izm} approaches Eq.~\eqref{eq:YukAlig} in an independent but complementary way compared to the Yukawa alignment. By the virtue of its conceptual transparency the models studied here are obtained. The singular alignment ansatz takes the singular value decomposition of Eq.~\eqref{eq:MassMatrix2HDM} as its starting point
\begin{equation} \label{eq:SVD}
    {\bf L}^\dagger {\bf M}_\text{diag} {\bf R} = v_1 {\bf Y}_1 + v_2 {\bf Y}_2 \;.
\end{equation}
Then, by noting that ${\bf M}_\text{diag} = \sum_i m_i {\bf P}_i$ with $[{\bf P}_i]_{jk} = \delta_{ij} \delta_{ik}$ ($i,j,k=1,2,3$) it is possible to redefine the l.h.s~to
\begin{equation}
    \sum_i m_i {\bf \Delta}_i = v_1 {\bf Y}_1 + v_2 {\bf Y}_2 \;,
\end{equation}
where ${\bf \Delta}_i = {\bf L}^\dagger {\bf P}_i {\bf R}$. In return, the ansatz means demanding that each Yukawa matrix satisfies
\begin{equation} \label{eq:SingAlig}
    {\bf Y}_k = \alpha_k {\bf \Delta}_1 + \beta_k {\bf \Delta}_2 + \gamma_k {\bf \Delta}_3 \;,\quad (k=1,2)\;.
\end{equation}
This gives place to the relations
\begin{equation}
\begin{gathered}
    m_1 = \sum_k \alpha_k v_k \;, \qquad
    m_2 = \sum_k \beta_k v_k \;,\\
    m_3 = \sum_k \gamma_k v_k \;.
\end{gathered}
\end{equation}
It is straightforward to see how substituting the ansatz in Eq.~\eqref{eq:SingAlig} into Eq.~\eqref{eq:SVD} guarantees the absence of tree-level FCNCs. Namely, each Yukawa matrix is diagonal in the mass-basis,
\begin{equation}
\bar{\bf{Y}}_k = {\bf L} \bf{Y}_k {\bf R}^\dagger =  \alpha_k {\bf P}_1 + \beta_k {\bf P}_2 + \gamma_k {\bf P}_3 \,.
\end{equation}
Moreover, when $\alpha_k \propto \beta_k \propto \gamma_k$ the ansatz implies the Yukawa alignment~\cite{Pich:2009sp}. In fact, singular alignment is equivalent to the generalized version of the Yukawa alignment~\cite{Penuelas:2017ikk} which is flavor non-universal. Interestingly, this kind of apparently \textit{ad hoc} ansatz could originate from a family symmetry as shown in Ref.~\cite{Varzielas:2011jr} or from an effective approach with additional hidden scalars~\cite{Serodio:2011hg}.
More importantly, the family universal alignment and its generalized non-universal version are both linear realizations of the minimal flavor violation (MFV) hypothesis~\cite{DAmbrosio:2002vsn}.
This implies that the appearance of FCNCs at loop-level poses no risk as all experimental bounds are respected, see for example~\cite{Buras:2010mh,Penuelas:2017ikk}.
Lastly, it has been shown recently that the family universal MFV ansatz can be generalized to a non-universal one~\cite{Egana-Ugrinovic:2018znw}; see~\cite{Egana-Ugrinovic:2019dqu} for an application to a 2HDM framework.

\begin{table}    \centering
    \begin{tabular}{lcc}
        Type- & A & B \\ \hline\hline
        $u_{3,R}$            & $\Phi_2$ & $\Phi_2$ \\
        $d_{3,R},\; e_{3,R}$ & $\Phi_1$ & $\Phi_2$ \\
        Other RH fermions    & $\Phi_1$ & $\Phi_1$
    \end{tabular}
    \caption{Each column shows the fermions with the same $\mathbb{Z}_2$ charge assignment as a certain Higgs doublet, $\Phi_{1(2)}$. This defines the new types A and B.
    Note that a flavor conserving ansatz is required in order to avoid tree-level FCNCs.}
    \label{tab:models}
\end{table}

\section{TWO NEW 2HDM TYPES}
\label{sec:models}
\noindent
We are now able to discuss two different models which we denote as Type-A and Type-B. The field content is that of a 2HDM with a softly-broken $\mathbb{Z}_2$ symmetry. The Type-A offers a collective distinction between the top quark and all the other fermions motivated by the big mass splitting. On the other hand, the Type-B creates a distinction between the whole third fermion family and the two light ones.
Only Type-A properly adopts the idea of natural small fermion masses, $m_f \ll m_t$, as they are all connected to the small VEV, $v_1$. 
In comparison, in Type-B the smallness of the bottom and tau masses cannot be called natural as $y_{b,\tau} \ll 1$. Despite this fact the scalar sector still possesses a natural small VEV. Thus, Type-B can be called natural.

The $\mathbb{Z}_2$ assignments for Type-A are:
\begin{align}
\begin{aligned}
    u_{3,R} &\to + u_{3,R} \\ 
    \{u_{j,R},d_{j,R},e_{j,R}\} &\to - \{u_{j,R},d_{j,R},e_{j,R}\}
\end{aligned}
\end{align}
and Type B:
\begin{align}
\begin{aligned}
    \{u_{3,R}, d_{3,R}, e_{3,R}\} &\to + \{u_{3,R}, d_{3,R}, e_{3,R}\} \\ 
    \{u_{j,R},d_{j,R},e_{j,R}\} &\to -\{u_{j,R},d_{j,R},e_{j,R}\}\;.
\end{aligned}
\end{align}
Here, $j$ denotes the remaining right-handed fermions. All left-handed ones are chosen even under the $\mathbb{Z}_2$ symmetry. We summarize the two models in Table~\ref{tab:models}.
The Yukawa Lagrangians are then expressed in Type-A as
\begin{align}
    -{\cal L}_Y^Q &=\sum_{i=1}^3
    \bar{Q}_{i,L}\! \left[ y_i^t \widetilde{\Phi}_2 u_{3,R} + \widetilde{\Phi}_1 (y_i^{c} u_{2,R} + y_i^{u} u_{1,R}) \right] \nonumber\\
    &+ \sum_{i=1}^3 \bar{Q}_{i,L} {\Phi}_1 (y_i^b  d_{3,R} +  y_i^{s} d_{2,R} + y_i^{d} d_{1,R}) + \text{h.c.} \nonumber\\
    -{\cal L}_Y^\ell &= \sum_{i=1}^3 \bar{\ell}_{i,L} {\Phi}_1 ( y_i^\tau  e_{3,R} +  y_i^{\mu} e_{2,R} + y_i^{e} e_{1,R}) + \text{h.c.}
    \label{eq:LyA}
\end{align}
and in Type-B as
\begin{align}
    -{\cal L}_Y^Q &= \sum_{i=1}^3
    \bar{Q}_{i,L}\! \left[ y_i^t \widetilde{\Phi}_2 u_{3,R} + \widetilde{\Phi}_1 (y_i^{c} u_{2,R} + y_i^{u} u_{1,R}) \right] \nonumber\\
    &+ \sum_{i=1}^3 \overline{Q}_{i,L}\! \left[ y_i^b {\Phi}_2 d_{3,R} + {\Phi}_1 ( y_i^s d_{2,R} + y_i^{d} d_{1,R}) \right] + \text{h.c.} \nonumber\\
    -{\cal L}_Y^\ell &= \sum_{i=1}^3 \bar{\ell}_{i,L} \!\left[ y_i^\tau {\Phi}_2 e_{3,R} + {\Phi}_1 (y_i^\mu e_{2,R} + y_i^{e} e_{1,R}) \right] + \text{h.c.} \;.
    \label{eq:LyB}
\end{align}
Generally speaking our two models feature tree-level FCNCs. However, as aforementioned discussed, through the introduction of the singular alignment ansatz we choose the right parameter region of family space such that the Yukawa matrices become diagonal in the mass basis. Thus, FCNCs are absent at tree-level. For further details we refer the reader to Ref.~\cite{Rodejohann:2019izm}.

Let us shortly notice an important feature of our two models related to the implied fermion mixing. As the fermion mass matrices are given in terms of two hierarchical VEVs, $v_1\ll v_2$, one can study the implications of setting the smaller one to zero. 
In Type-A all mass matrices are equal to zero except the one for the up-type quarks which takes the form
\begin{equation} \label{eq:vevless}
    {\bf M}_u = v_2 \begin{pmatrix}
    0 & 0 & y_1^t \\
    0 & 0 & y_2^t \\
    0 & 0 & y_3^t
    \end{pmatrix} \;.
\end{equation}
As the down-type quarks have a null mass matrix a simultaneous unitary transformation in the quark weak doublet leaves the kinetic terms invariant, and simultaneously brings us to the mass basis. Therefore, at this level the quark mixing matrix is given by the identity which is a good first approximation to the observed quark mixing matrix.

For the sake of completeness, and in order to discuss also lepton mixing one must introduce massive neutrinos. Let's assume we have done that without specifying them to be of Dirac or Majorana nature. Under this circumstance, as the mass matrices for both the charged leptons and neutrinos depend on the same VEV (even in the Majorana scenario), their mixing is expected to strongly deviate from the identity and behave more anarchically~\cite{Hall:1999sn} which is a good description of observations.

On the other hand, for Type-B all mass matrices take the form of Eq.~\eqref{eq:vevless} in the limit $v_1 \to 0$. In return, this implies that all fermion should mix anarchically which is not the case. This undesired issue can be solved by reassigning all $\mathbb{Z}_2$ odd charges to the left-handed fermions instead of the right-handed ones. Thereafter, a weak-basis transformation in the right-handed fields would be enough to diagonalize the Yukawa matrices and recover the trivial quark mixing. However, in the lepton sector one would have anarchic mixing only if Majorana neutrinos are assumed.

Hence, in terms of fermion mixing both models are able to predict trivial mixing for the quark sector (under the right $\mathbb{Z}_2$ charge assignment) and anarchic mixing for the lepton sector (if neutrinos are considered as Majorana particles). Fermion mixing has been explicitly related to $t_\beta$ in the recent study in Ref.~\cite{CarcamoHernandez:2020ney}. Moreover, similar conclusions were obtained for Type-B in~\cite{Botella:2016krk}.

The Yukawa Lagrangian in the mass basis is expressed by:
\begin{align}
    -{\cal L}_Y & \supset \sum_{f} \frac{m_f}{(246 \,\text{GeV})} \Big( \xi^h_f \bar{f}fh +\xi^H_f \bar{f}fH -i\xi^A_f \bar{f}\gamma_5 f A \Big) \nonumber \\
    & -H^+\frac{\sqrt{2}\sum_{ij} {\bf V}_{ij}^\text{CKM}}{(246 \,\text{GeV})}\ \bar{u}_i\left( m_{u_i}\xi^{H^+}_{q_u}P_L + m_{d_j}\xi^{H^+}_{q_d}P_R \right) d_j \nonumber \\
    & -H^+ \frac{\sqrt{2}m_\ell}{(246 \,\text{GeV})}\ \xi^{H^+}_\ell \bar{\nu}_{L,i} \ell_{R,j} +  \text{h.c.}
    \label{Eq.MasL}
\end{align}
where ${\bf V}^\text{CKM}$ is the quark mixing matrix. 
The SM is recovered for $\xi^h_f = 1$ and $\xi^{H,A,H^+}_f \!= 0$. In Table~\ref{tab:couplings1} we show the corresponding couplings for the conventional NFC scenarios, while in Table~\ref{tab:couplings2} the respective ones for our Types-A and B. The two tables show great similarities, as the main change from the conventional ones is breaking their family universality.

To derive the Yukawa couplings shown in Tables~\ref{tab:couplings1} and~\ref{tab:couplings2} we insert $\Phi_{1,2}$ from Eq.~\eqref{eq:scalar} into Eqs.~\eqref{eq:LyA} and~\eqref{eq:LyB}.
We change to the mass basis by performing a rotation in the neutral and charged scalar sector as in Eq.~\eqref{eq:rotation}. The resulting terms depend on $\beta$ and $\alpha$ as well as on the two VEVs, $v_{1,2}$.
In addition, we use the relations between the fermionic Yukawa couplings and masses,
\begin{align}
   y_f = \frac{m_f}{c_\beta\, v} \qquad \text{or} \qquad y_f = \frac{m_f}{s_\beta\,v} \;.
\end{align}
Here, the former relation should be used if the given fermion couples to $\Phi_1$ and the latter if it couples to $\Phi_2$.
After carefully following these steps we arrive at the couplings shown in Tables~\ref{tab:couplings1} and~\ref{tab:couplings2}.
See Ref.~\cite{Chulia:2019evx} for an example where the couplings acquire a completely different behaviour when enlarging the flavor symmetry to a larger group. 
For other related phenomenological applications of the Yukawa alignment see for example~\cite{Li:2020dbg}.

\begin{table}[t]    \centering
\begin{tabular}{ccccc}
  Type- & I & II & X & Y \\
  \hline \hline
  $\xi^h_{q_u}$  & $c_\alpha/s_\beta$ & $c_\alpha/s_\beta$ & $c_\alpha/s_\beta$ & $c_\alpha/s_\beta$ \\
  $\xi^h_{q_d}$ & $c_\alpha/s_\beta$ & $-s_\alpha/c_\beta$ & $c_\alpha/s_\beta$ & $-s_\alpha/c_\beta$ \\
  $\xi^h_{\ell}$ & $c_\alpha/s_\beta$ & $-s_\alpha/c_\beta$ & $-s_\alpha/c_\beta$ & $c_\alpha/s_\beta$ \\
  \hline
  $\xi^H_{q_u}$  & $s_\alpha/s_\beta$ & $s_\alpha/s_\beta$ & $s_\alpha/s_\beta$ & $s_\alpha/s_\beta$ \\
  $\xi^H_{q_d}$ & $s_\alpha/s_\beta$ & $c_\alpha/c_\beta$ & $s_\alpha/s_\beta$ & $c_\alpha/c_\beta$ \\ 
  $\xi^H_{\ell}$ & $s_\alpha/s_\beta$ & $c_\alpha/c_\beta$ & $c_\alpha/c_\beta$ & $s_\alpha/s_\beta$ \\
  \hline
  $\xi^A_{q_u}$  & $1/t_\beta$ & $1/t_\beta$ & $1/t_\beta$ & $1/t_\beta$ \\
  $\xi^A_{q_d}$  & $-1/t_\beta$ & $t_\beta$ & $-1/t_\beta$ & $t_\beta$ \\
  $\xi^A_{\ell}$ & $-1/t_\beta$ & $t_\beta$ & $t_\beta$ & $-1/t_\beta$ \\
  \hline
  $\xi^{H^+}_{q_u}$  & $1/t_\beta$ & $1/t_\beta$ & $1/t_\beta$ & $1/t_\beta$ \\
  $\xi^{H^+}_{q_d}$ & $1/t_\beta$ & $-t_\beta$ & $1/t_\beta$ & $-t_\beta$ \\ 
  $\xi^{H^+}_{\ell}$ & $1/t_\beta$ & $-t_\beta$ & $-t_\beta$ & $1/t_\beta$ \\
\end{tabular}
\caption{Flavor universal Yukawa couplings of the charged fermions to the Higgs bosons $h$, $H$, $A$, and $H^+$ in the four conventional 2HDMs.} 
\label{tab:couplings1}
\end{table}

The couplings of the CP-even scalars, $h$ and $H$, to a pair of vector bosons, $V=W^\pm,Z$, are modified by
\begin{equation}
    \xi_{VV}^h = s_{\beta-\alpha} \quad \text{and} \quad
    \xi_{VV}^H = c_{\beta-\alpha} \;.
\end{equation}
The SM values are favored by present data. This means that to a very good degree of approximation,
\begin{equation}
    \sin (\beta - \alpha) \simeq 1 \;.
\end{equation}
This is called the alignment limit (AL).
It defines the condition for $h$ to be SM-like besides the correct mass.
In terms of angles we can approach this limit as: $\beta = \alpha + \pi/2 -\epsilon$ with $\epsilon \to 0$. 
As an implication of this one obtains $\xi_f^h \to 1$ in Tables~\ref{tab:couplings1} and~\ref{tab:couplings2}.
This again shows the fact that $h$ behaves as the SM Higgs in the AL. Therefore, our two proposed models satisfy the same alignment conditions as the conventional NFC ones.
To better understand the behaviour of the Yukawa couplings we rewrite the relevant $\xi^{(h,H)}_f$ in terms of $t_\beta$, $c_{\beta-\alpha}$, and $s_{\beta-\alpha}$.
Including terms for the AL up to $\mathcal{O}(\epsilon)$ leads to
\begin{align}
\begin{aligned}
    {c_\alpha}/{s_\beta} &= s_{\beta -\alpha} + c_{\beta -\alpha}/t_\beta \, \simeq 1 +\epsilon/ t_\beta \; ,\\
    -{s_\alpha}/{c_\beta}&= s_{\beta -\alpha} - c_{\beta -\alpha}\ t_\beta \; \simeq 1 -\epsilon \, t_\beta \;,\\
    {c_\alpha}/{c_\beta} &= c_{\beta -\alpha} + s_{\beta -\alpha}\ t_\beta \; \simeq \epsilon + t_\beta \;,\\
    {s_\alpha}/{s_\beta} &= c_{\beta -\alpha} - s_{\beta -\alpha}/t_\beta \, \simeq \epsilon - 1/ t_\beta \;.
\end{aligned}
\end{align}
Thus, away from the exact AL significant deviations are expected in Type-A and B compared to the SM Higgs couplings to first and second generation fermions.

\begin{table}[t]    \centering
\begin{tabular}{ccc}
  Type- & A & B \\
  \hline\hline
  $\xi^h_t$ & $c_\alpha/s_\beta$ & $c_\alpha/s_\beta$ \\
  $\xi^h_{b,\tau}$ & $-s_\alpha/c_\beta$ & $c_\alpha/s_\beta$ \\
  $\xi^h_\text{light}$ & $-s_\alpha/c_\beta$ & $-s_\alpha/c_\beta$ \\
  \hline
  $\xi^H_t$ & $s_\alpha/s_\beta$ & $s_\alpha/s_\beta$ \\
  $\xi^H_{b,\tau}$ & $c_\alpha/c_\beta$ & $s_\alpha/s_\beta$ \\
  $\xi^H_\text{light}$ & $c_\alpha/c_\beta$ & $c_\alpha/c_\beta$ \\
  \hline
  $\xi^A_t$ & $1/t_\beta$ & $1/t_\beta$ \\
  $\xi^A_{b,\tau}$ & $t_\beta$ & $-1/t_\beta$ \\
  $\xi^A_{u,c}$ & $-t_\beta$ & $-t_\beta$ \\
  $\xi^A_{d,s,\ell}$ & $t_\beta$ & $t_\beta$ \\
  \hline
  $\xi^{H^\pm}_t$ & $1/t_\beta$ & $1/t_\beta$ \\
  $\xi^{H^\pm}_{b,\tau}$ & $-t_\beta$  & $1/t_\beta$ \\
  $\xi^{H^\pm}_\text{light}$ & $-t_\beta$ & $-t_\beta$
\end{tabular}
\caption{Flavor non-universal Yukawa couplings, cf.~Eq.~\eqref{Eq.MasL}, of the charged fermions to the scalars $h,\, H,\, A$, and $H^+$ in Type-A and B with $\text{light}=\{u,c,d,s,\ell\}$ and $\ell =\{e,\mu\}$.}
\label{tab:couplings2}
\end{table}

We note that regarding the Higgs couplings Type-A is closely related to Type-II besides the up and charm quark couplings. Due to their small Yukawa values they have limited phenomenological relevance at colliders, although some efforts have been made to constraint their values. In particular, the decay $h\to J/\Psi +\gamma$ is sensitive to potential deviations in the charm-Yukawa coupling which could be test. We investigate this in the next section.

For Type-B the situation is slightly different. Compared to Type-I the couplings to $d,\, s,\, u,\, c,\, e, \mu$ are changed. Therefore, in Type-B in case of deviations from the AL those couplings can be enhanced for large values of $t_\beta$ instead of being suppressed as in Type-I.
The changes in the muon coupling are of special interest as it is experimentally constrained~\cite{Sirunyan:2018koj, Aad:2019mbh,Sirunyan:2019tkw}.
We further investigate this in the next sections. For last, Type-I is mostly constrained for $t_\beta \lesssim 10$ due to the $1/t_\beta$ suppressed $b$-Yukawa coupling, while for Type-II relevant constraints also arise for large values of $t_\beta$~\cite{Kling:2020hmi}.

\begin{figure}[t]
    \includegraphics[width=\linewidth]{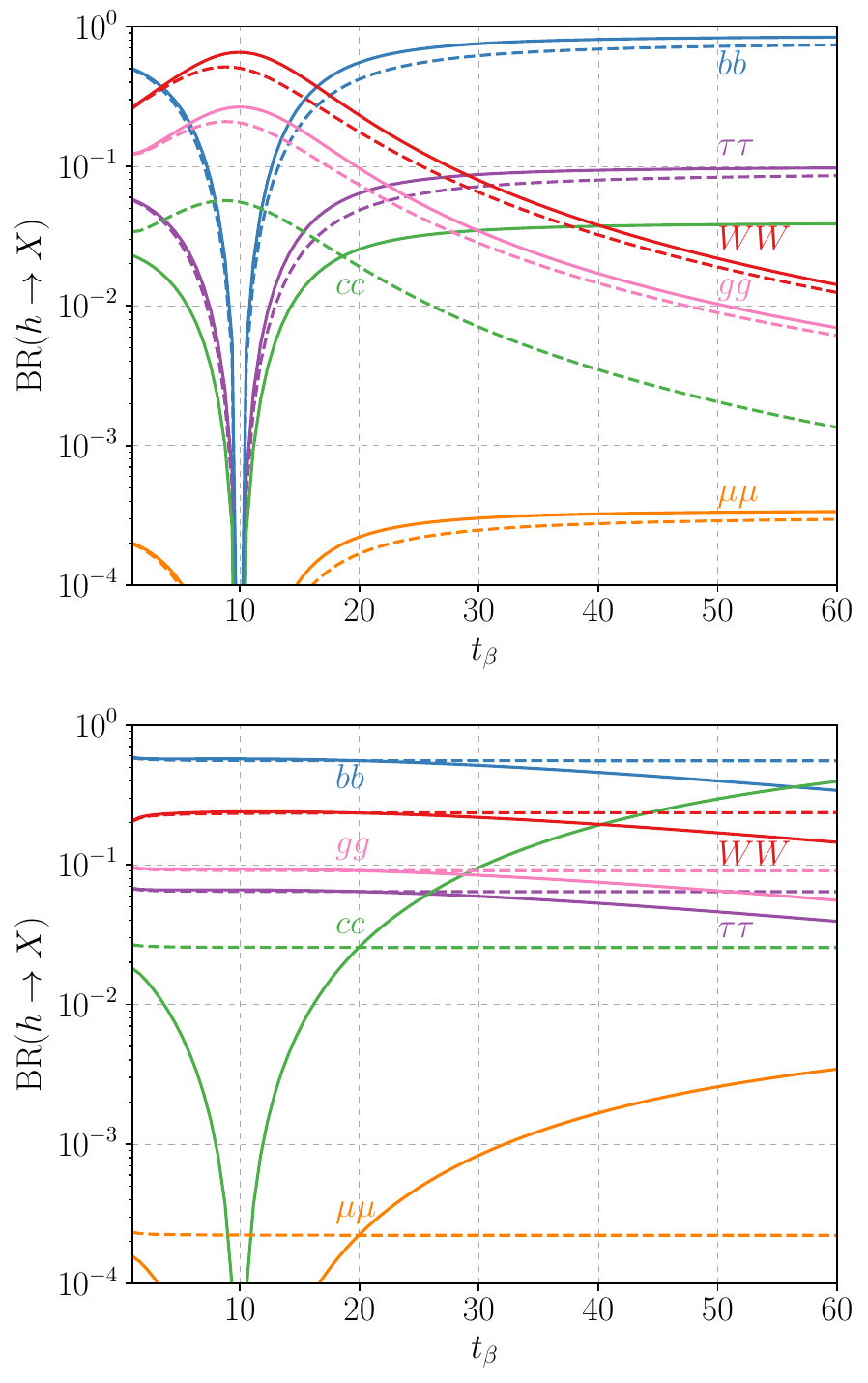}
     \caption{Branching ratios of the SM-like scalar $h$ in Type-A (top) and B (bottom) for $c_{\beta-\alpha}=0.1$ \cite{Djouadi:2005gi,Djouadi:2005gj}. For comparison we show the BRs in Type II (top) and I (bottom) as dashed lines using the same color for each mode.}
     \label{fig:BR_h}
\end{figure}

\begin{figure}[t]
	\includegraphics[width=\linewidth]{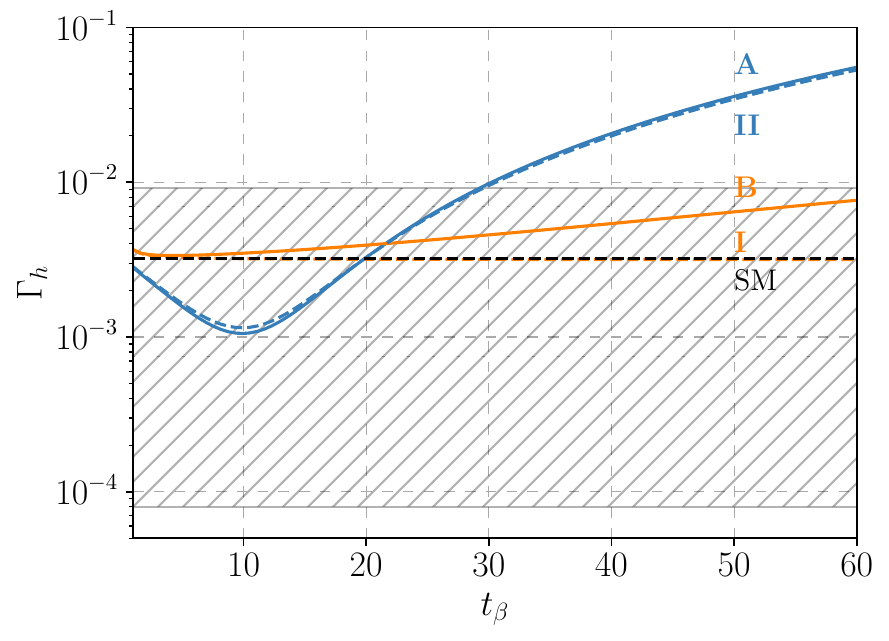}
	\includegraphics[width=\linewidth]{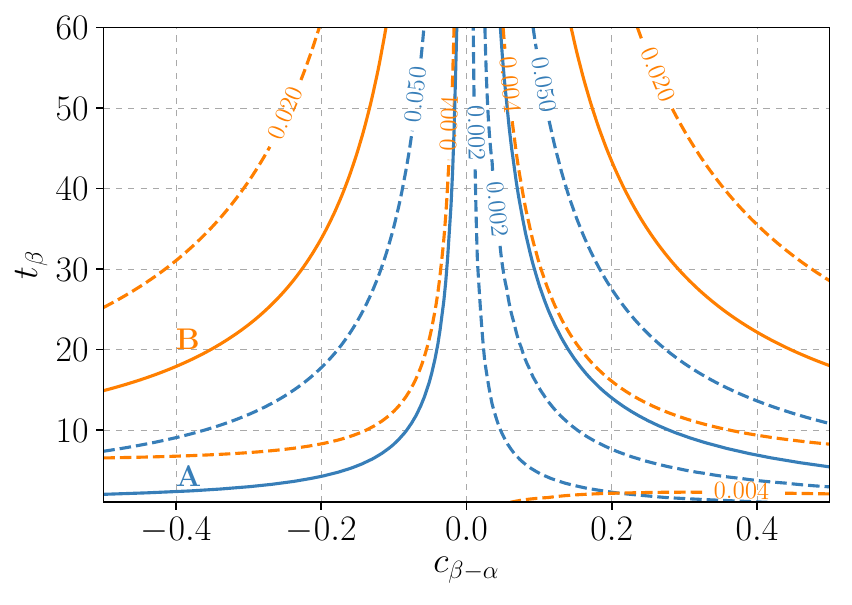}
	\caption{Top pannel: Total decay width of $h$ for Type-A, B, I, and II with $c_{\beta-\alpha}=0.1$. The SM value is depicted as the black dotted line and the experimentally allowed band~\cite{Sirunyan:2019twz} at 95$\%$ CL\,as the grey hatched region.\\
	Lower panel: Contours of $\Gamma_h=$const.\,for Type-A (blue) and B (orange). Regions outside the solid lines are excluded at 95$\%$ CL~\cite{Sirunyan:2019twz}.}
	\label{fig:TotalWidth}
\end{figure}

To visualize deviations from the SM Higgs couplings and the differences of the four types we show the branching ratios (BRs) of $h$ as a function of $t_\beta$ in Figure~\ref{fig:BR_h} with $c_{\beta -\alpha}=0.1$ for Type-A and II (top), as well as for Type-B and I (bottom)~\cite{Djouadi:2005gi,Djouadi:2005gj}.
While for Type-A most decay modes behave very similar to Type-II the BR$(h\to c\bar c)$ differs significantly. In Type-B all BRs show distinct behavior compared to Type-I for $t_\beta \gtrsim 30$ as BR$(h\to c\bar c)$ becomes sizeable. Similarly, BR$(h\to \mu^+\mu^-)$ also shows deviations from predictions in the usual 2HDM-types. A detailed discussion of this mode and the Higgs decay into charmonium plus a photon is presented in the next section. In addition, the total decay width deviates stronger from the SM value for Type-A than for Type-B as shown in Figure~\ref{fig:TotalWidth}.
In the lower panel regions outside the solid contour lines are excluded at 95$\%$ CL by CMS~\cite{Sirunyan:2019twz}.

\begin{figure}[t]
     \includegraphics[width=\linewidth]{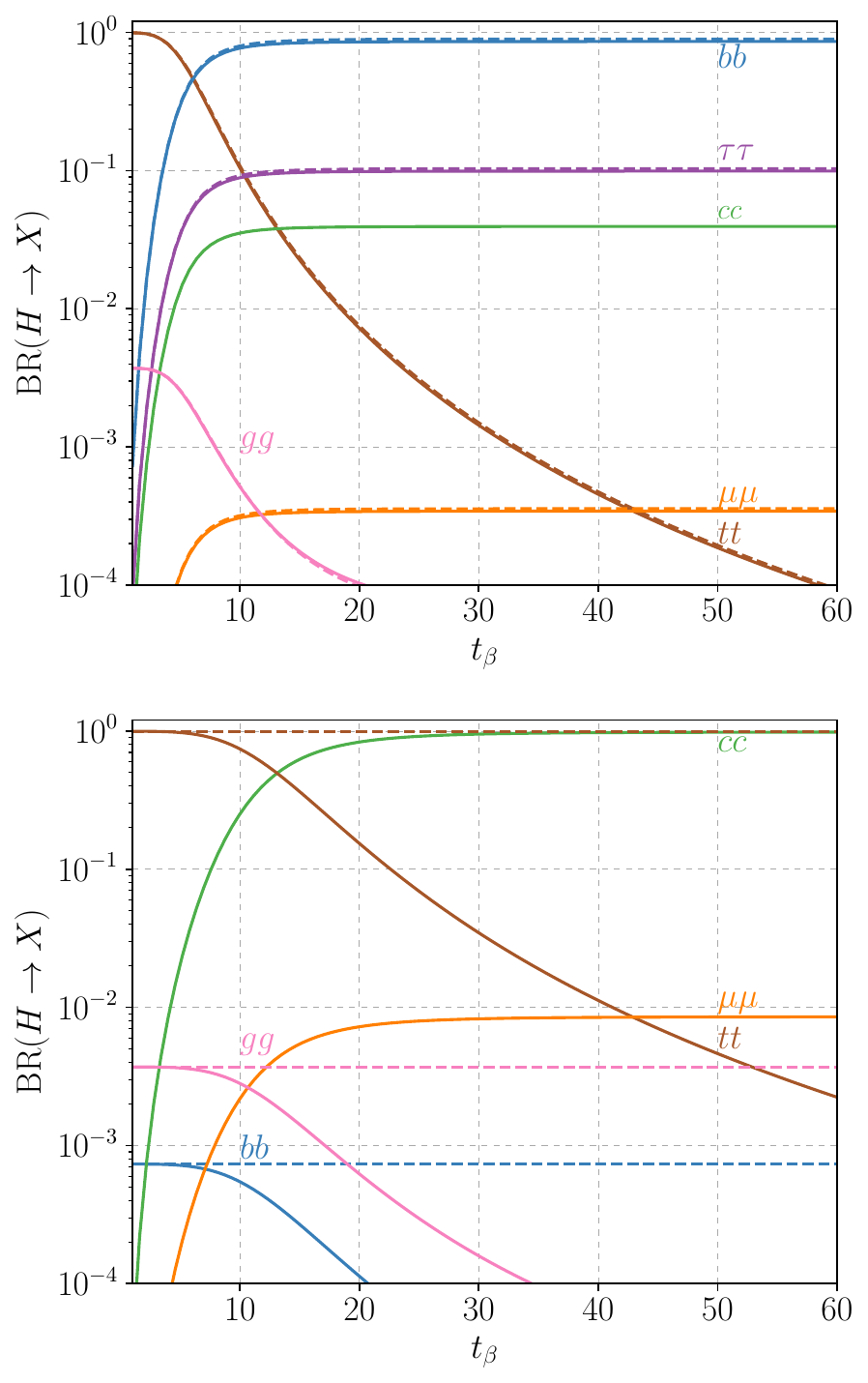}
     \caption{Dominant BRs of the heavy CP-even scalar, $H$, in Type-A (top) and B (bottom) for $c_{\beta-\alpha}=0$ and $m_H=500$\,GeV. For comparison we also show the BRs for Type-II (top) and Type-I (bottom) as dashed lines. The BRs of the pseudoscalar, $A$, behave similarly.}
     \label{fig:BR_H}
\end{figure}

In a similar manner we show the BRs for the heavy scalar, $H$, in Figure~\ref{fig:BR_H}. The BRs of the pseudoscalar, $A$, behave similarly. Therefore, we do not discuss them explicitly.
Here, BR$(H\to t \bar{t})$ dominates for $t_\beta \lesssim 5$ in Type-A and II, respectively for $t_\beta \lesssim 12$ in Type-B, and for all values of $t_\beta$ in Type-I.
For values of $t_\beta\gtrsim 10$ BR$(H\to b \bar{b})$ becomes dominant in Type-A and II, while in Type-B the BR$(H\to c \bar{c})$ takes over.
In Type-A and II the decay $H\to \tau^+ \tau^-$ features the second biggest BR for $t_\beta\gtrsim 10$. Only in Type-A the decay $H\to c\bar{c}$ reaches a relevant BR.
In Type-B the BR($H\to \mu^+ \mu^-$) reaches a significant value for $t_\beta \gtrsim 10$.
In Type-I the ratios of the BRs stay constant and the $gg$ channel is next-to-dominant.
Considering deviations from the AL decays to weak gauge bosons become relevant. For example for $c_{\beta-\alpha}=0.1$ the maxima of BR$(H\to WW)\simeq 0.33$ and BR$(H\to ZZ)\simeq 0.16$ are at $t_\beta \approx 5$. For higher values of $t_\beta$ both approach values of $\mathcal{O}(1\%)$.

Recently, the importance of studying the couplings of the SM Higgs as a way to distinguish among different multi-scalar scenarios was highlighted in Ref.~\cite{Arroyo-Urena:2020fkt}. In particular, different imprints that the SM and BSM scenarios leave on the Higgs Yukawa couplings were identified. Namely, in the SM those couplings lay on a single line if plotted as a function of the fermion mass. This also occurs for the 2HDM of Type-I but with a different slope. On the other hand, for the Type-II the Higgs Yukawa couplings will lay on two lines, one for down-type quarks and leptons and one for up-type quarks. Now, within Type-A the top Yukawa coupling will deviate from the line defined by the remaining fermions. In Type-B all third generation Yukawa couplings will lay on a different line than the ones of the light fermions.

\section{MODEL CONSTRAINTS AND BENCHMARKS}  \label{sec:pheno}
\noindent

There are mainly two ways to study multi-scalar scenarios: i) by possible deviations from the SM predictions, like Higgs properties, and ii) by direct searches for the new scalar states. For our analysis we take both into consideration.

While a full investigation of the 2HDM parameter space is beyond the scope of this article, we make use of results derived for the well-studied Types-I and II, and discuss differences occurring in our Types-A and B.
In particular, we refer the reader to Ref.~\cite{Kling:2020hmi}. There a summary of the most relevant constraints and a comprehensive analysis regarding the present situation of the 2HDM neutral scalars from current LHC searches is presented. 

\subsection{CONSTRAINTS}
\label{sec:constr}
\noindent
We consider the following experimental constraints for 2HDMs as the most relevant ones for our study:
\begin{itemize}
    \item {\bf Higgs couplings}: As aforementioned the couplings of the SM-like Higgs, $h$, get modified. 
    The Higgs coupling modifiers, $\kappa_i$, are defined as the ratio of the Higgs coupling to the corresponding SM value $\kappa_i= g_{hii}/g^\text{SM}_{hii}$, where $i$ denotes a SM field~\cite{Sirunyan:2018koj,Aad:2019mbh}. This in turn implies
    \begin{equation}
        \kappa_i^2 = \frac{\sigma_i}{\sigma_i^\text{SM}} \quad\text{ or }\quad 
        \kappa_i^2 = \frac{\Gamma_i}{\Gamma_i^\text{SM}} \,,
    \end{equation}
    which correspond to $(\xi_i^h)^2$ in our models. In Table~\ref{tab:kappas} we summarize the current limits on those derived from combined measurements of ATLAS and CMS~\cite{Sirunyan:2018koj,Aad:2019mbh}.
    Of special interest for the models under consideration are the channels $h\to\mu\mu$ and $h\to J/\Psi+\gamma$. In those decays the coupling structure differs from the conventional types.

    In addition, the recent limit on the total decay width of the SM Higgs by CMS~\cite{Sirunyan:2019twz} ($0.08\,\text{MeV}<\Gamma_h<9.16$\,MeV at $95\,\%$ CL) strongly constrains enhanced couplings of $h$ to fermions and additional decay modes. In this regard, Figure~\ref{fig:TotalWidth} shows the observed limit together with $\Gamma_h$ as a function of $t_\beta$ in Types-I, II, A, and B for $c_{\beta-\alpha}=0.1$ as well as a contour plot in the $(c_{\beta-\alpha},t_\beta)$ plane.

    \begin{table}    \centering 
    \begin{tabular}{lccc}
     & Bayesian fit~\cite{deBlas:2018tjm} & CMS~\cite{Sirunyan:2018koj} & ATLAS~\cite{Aad:2019mbh} \\
    \hline\hline
        $\kappa_W$ & $1.01\pm 0.06$ & $1.10^{+0.12}_{-0.17}$ & $1.05\pm0.08$ \\
        $\kappa_Z$ & $1.01\pm 0.06$ & $0.99^{+0.11}_{-0.12}$ & $1.10\pm 0.08$ \\
        $\kappa_t$ & $1.04^{+0.09}_{-0.10}$ & $1.11^{+0.12}_{-0.10}$ & $1.02^{+0.11}_{-0.10}$ \\
        $\kappa_b$ & $0.94\pm 0.13$ & $-1.10^{+0.33}_{-0.23}$ & $1.06^{+0.19}_{-0.18}$\\
        $\kappa_\tau$ & $1.0 \pm 0.1$ & $1.01^{+0.16}_{-0.20}$ & $1.07\pm0.15$ \\
        $\kappa_\mu$ & $0.58^{+0.40}_{-0.38}$ & $0.79^{+0.58}_{-0.79}$ &  $< 1.53$ at 95\%\,C.L. \\
    \end{tabular}
    \caption{Current 68\% probability sensitivities and best fit values for the Higgs coupling modifiers, $\kappa_i$, as obtained from a Bayesian statistical analysis and from combined data taken by ATLAS and CMS at $\sqrt{s} = 13$ TeV. The ATLAS fit assumes all coupling modifiers to be positive.}
    \label{tab:kappas}
    \end{table}

    \item {\bf Direct collider searches}: 
    As an example we explicitly consider the ATLAS and CMS searches for heavy scalar resonances decaying to muon pairs~\cite{Aaboud:2019sgt, Sirunyan:2019tkw}. In those searches we expect significant deviations for Type-B compared to Type-I. In~\cite{Sirunyan:2019tkw} model-independent exclusion limits on the production cross section times the BR have been determined for scalars in the mass range from 130 to 1000\,GeV.
    In other channels the behavior is expected to be similar to the conventional types. As shown in Ref.~\cite{Kling:2020hmi} values of $t_\beta \gtrsim 10$ are excluded by searches for $A/H \to \tau^-\tau^+$ for the mass degenerated scenario and $c_{\beta-\alpha}=0.05$. Therefore, we adapt these constraints for Type-II and A.

    \item {\bf Electroweak Precision Constraints}: The two terms in the scalar potential proportional to $\lambda_4$ and $\lambda_{5}$ break the custodial symmetry. This leads to additional contributions to the $\rho$ parameter which can be avoided by taking $m_A= m_{H^\pm}$ or/and $m_H = m_{H^\pm}$ \cite{Gerard:2007kn,deVisscher:2009zb,Grzadkowski:2010dj}.

    \item {\bf Flavor Observables}: Even after avoiding FCNCs at tree-level in 2HDMs they can arise at loop-level from charged Higgs loops. Constraints from the Belle II dataset~\cite{Misiak:2017bgg, Belle:2016ufb} on $b\to s\gamma$ decays are especially relevant. They require $m_{H^\pm}> 600$\,GeV. The Type-II is most sensitive to this constraint. After a careful study of the involved couplings it is possible to show that our two proposed types behave in the same way as the Type-II for this flavor violating transition. Therefore, we require in the following that the charged scalar mass lies above $600\,\text{GeV}$. We note that these flavor constraints are model dependent and could be relaxed in the presence of more intricate BSM sectors.
\end{itemize}

\subsection{BENCHMARK SCENARIOS}
\label{sec:benchmark}
\noindent
The number of independent free parameters in the scalar sector is seven. We choose them to be given by
\begin{equation}
    \{m_{12}^2,\, m_h,\, m_H,\, m_A,\, m_{H^\pm},\, t_\beta,\, \alpha \} \;.
\end{equation}
To simplify the analysis and reduce the number of free parameters we investigate the most relevant phenomenological aspects of our two models by focusing on the following well motivated benchmarks:
\begin{itemize}
    \item Alignment limit: Two facts may help us to reduce the parameter space. We have $\beta = \alpha + \pi/2$. In addition, we know that $v_2 \gg v_1$ such that we could take $v_1 \in (3, 58) \,\text{GeV}$. Here, the lower bound is obtained by demanding the bottom Yukawa coupling to be of order one, $y_b\sim {\cal O}(1)$, i.e.~$m_b\approx 3$~GeV. The upper bound is obtained by relaxing the previous condition and just demanding $v_1 /v_2 \lesssim {\cal O}(10^{-1})$. In return, we obtain a region for $t_\beta \in (3,58)$ that implies for the scalar mixing angle
    \begin{equation}
        \alpha \in (-18.43,-0.99)^\circ \;.
    \end{equation}
    In the AL flavor universality in the Yukawa couplings is restored for $h$ but not for $H$ and $A$.

    We recall the two employed criteria for naturalness, as they are crucial to understand why we conceive $t_\beta \in (3,\,58)$ as the natural range for our discussion. In the Type-A only for $t_\beta \sim (20,\,58)$ hierarchical fermion masses ($m_b,m_\tau,m_c$) are natural, whereas for $t_\beta \sim (3,20)$ they stop being so. However, we employ the latter range as it is connected to a \textit{natural} small value of $v_1$. For Type-B there is no range of natural hierarchical fermion masses (except $t_\beta \gg 58$).

    \item Degenerate masses: Contributions to the oblique parameter $T$ (or $\rho$) are the most restrained ones. As they basically depend on the relative mass squared differences, one may define three different cases: i) $m_H = m_{H^\pm}$, ii) $m_A = m_{H^\pm}$, and iii) $m_H = m_A = m_{H^\pm}$. The bounds from electroweak precision measurements can be more easily satisfied in the last case.

    \item Unitarity and vacuum stability: It has been shown that in the AL the soft $\mathbb{Z}_2$ breaking parameter is fixed to
    \begin{equation} \label{eq:m12constraint}
        m_{12}^2 = \frac{t_\beta}{1+ t^2_\beta}\ m_H^2 \;,
    \end{equation}
    by unitarity and vacuum stability \cite{Kling:2016opi}.
    Nonetheless, away from the AL and for large $t_\beta$ perturbativity requires the soft-breaking parameter to satisfy~\cite{Kling:2016opi}
    \begin{equation}
       | m_{12}^2 - m_{H}^2\, s_\beta c_\beta | \lesssim v^2 \;.
    \end{equation}
\end{itemize}

The simultaneous employment of the different benchmark scenarios significantly reduces the number of parameters. In its two minimal forms the analysis could require three or four free parameters. This is a consequence of Eq.~\eqref{eq:m12constraint}, the AL, and the degenerate masses assumption.
In the following, we will employ these benchmarks as a complimentary aspect of our discussions.

\begin{figure}[t]
	\includegraphics[width=\linewidth]{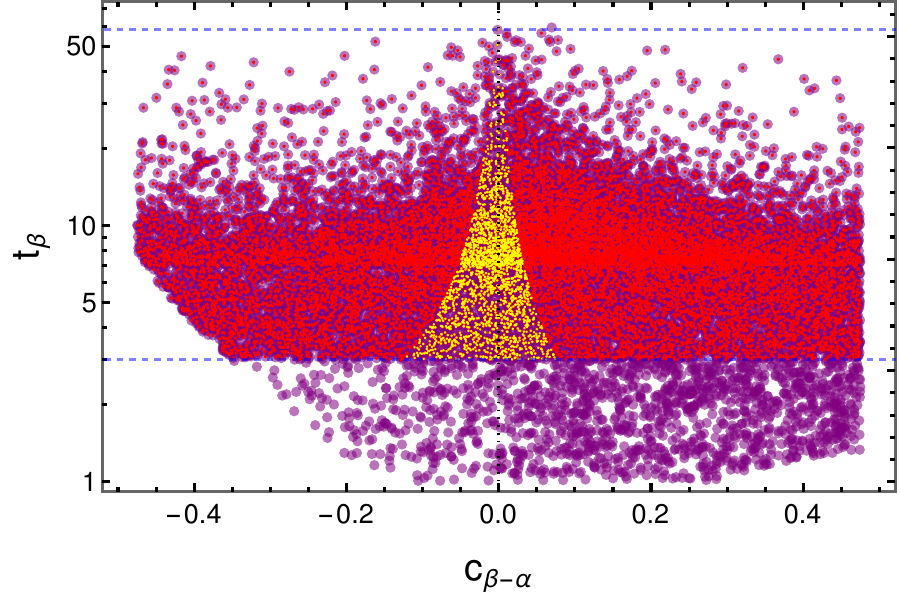}
	\caption{Current allowed regions from the measured SM-like Higgs couplings to fermions, $\kappa^h_{t,b,\tau}$, and gauge bosons, $\kappa_V^h$, at 95\% CL for 2HDMs of Type-I (purple), B (red), A and II (both in yellow). The vertical dashed line corresponds to the AL, whereas the two horizontal ones to the previously discussed \textit{natural} range ($3< t_\beta < 58$).}
	\label{fig:GeneralCase}
\end{figure}

\section{PHENOMENOLOGICAL RESULTS}
\label{sec:results}
\begin{figure}[t]
	\includegraphics[width=\linewidth]{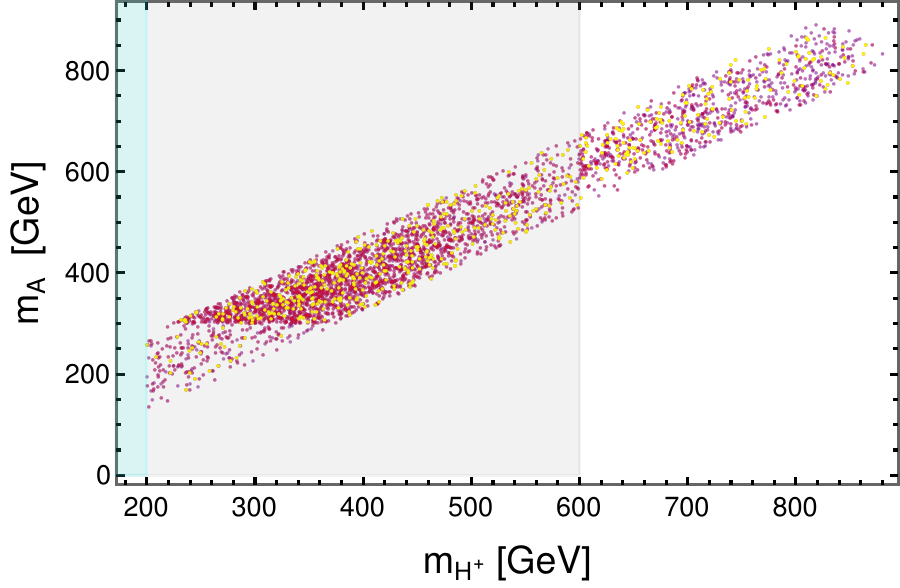}
	\caption{Allowed masses for the 2HDM Type-I (purple), B (red), A and II (both in yellow) in the bi-dimensional plane ($m_{H^\pm}$, $m_A$). The gray (applicable to Type-II, A, and B) and cyan (applicable to Type-I and $t_\beta < 2$) regions are excluded by flavor constraints (mostly $b\to s$ transitions). A similar looking plot can be obtained for ($m_{H^\pm}$, $m_H$).}
	\label{fig:mAvsmHpGC}
\end{figure}

\noindent
We start this section by commenting on the plane spanned by $t_\beta$ and $c_{\beta -\alpha}$. In Figure~\ref{fig:GeneralCase} we show the allowed regions from the measured SM-like Higgs couplings shown in Table~\ref{tab:kappas}.
Each plotted point satisfies the contributions to the oblique parameter $T$, BFB, unitarity, perturbativity, and global minimum conditions. Also, the charged scalar mass is required to be $m_{H^\pm} > 600$\,GeV as implied by the $b \to s$ flavor violating transitions. The range of scalar masses satisfying the previous conditions are shown in Figure~\ref{fig:mAvsmHpGC}. In both figures the two new types (A and B) can be compared with the conventional ones (I and II).

\begin{figure}[t]
    \includegraphics[width=\linewidth]{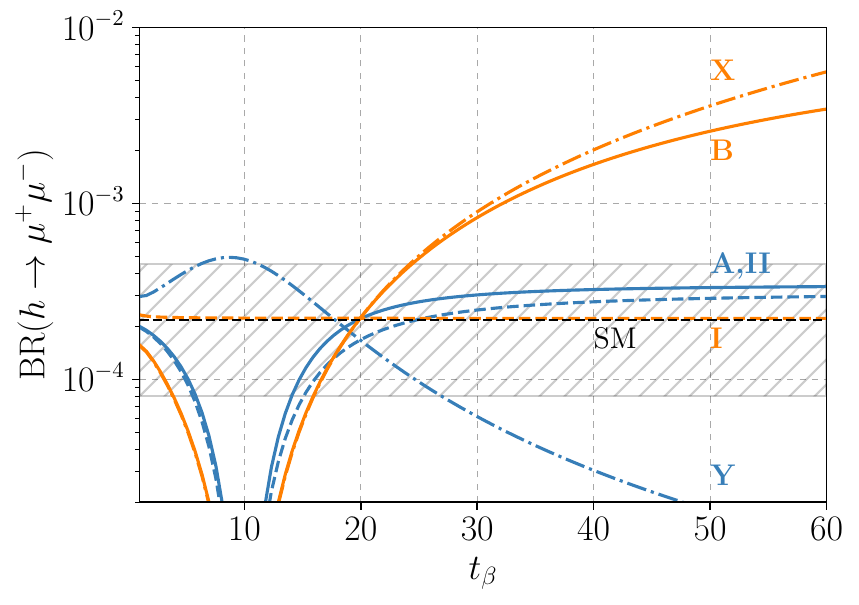}
    \caption{BR$(h \to \mu^+\mu^-)$ in all six 2HDM types for $c_{\beta-\alpha}=0.1$. The SM value is shown as a dashed black line and the experimental allowed region at $95\%$ CL as the hashed band~\cite{CMS:2020eni}.}
    \label{fig:BR_mumu}
\end{figure}

One of the salient features of our models concerns the Higgs coupling with muons which could significantly deviate from the SM. Recently the CMS collaboration announced results for the Higgs decay into a muon pair. The obtained limits are $0.8 \times 10^{-4} < \text{BR}(h \to \mu^+\mu^-) < 4.5  \times 10^{-4}$ at $95\%$ CL~\cite{CMS:2020eni}. A comparison of BR$(h\to\mu^+\mu^-)$ in the four common types (I, II, X, Y) and our types (A, B) is depicted in Figure~\ref{fig:BR_mumu} for $c_{\beta-\alpha}=0.1$. We note that large values of $t_\beta$ are excluded for Type-B, while for Type-A most of the range of $t_\beta$ is still consistent with the data.

\begin{figure}[t]
    \includegraphics[width=\linewidth]{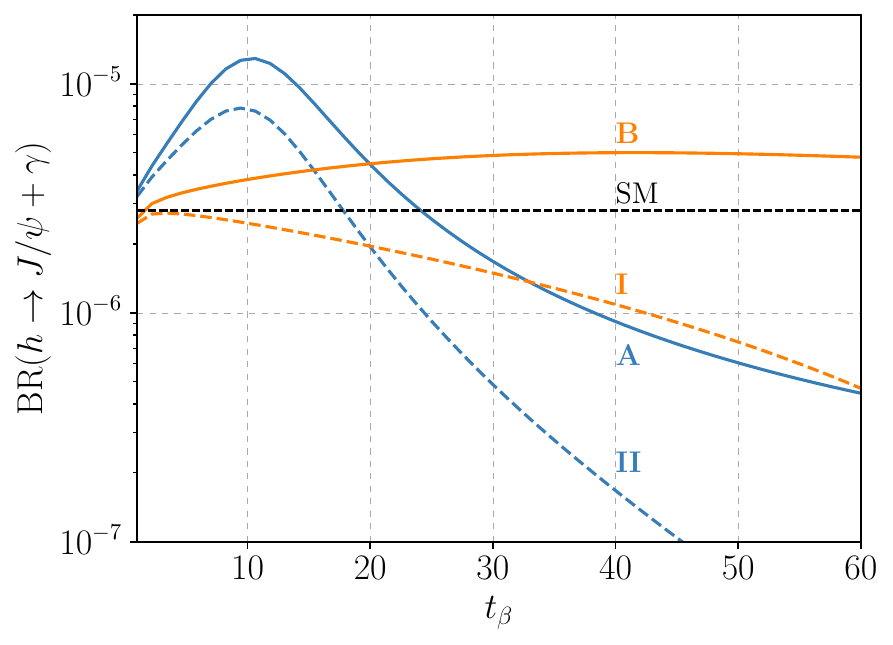}
    \caption{BR$(h \to J/\psi+\gamma)$ in the four 2HDMs types (Type-X and Y are identical to Type-I and II, respectively) for $c_{\beta-\alpha}=0.1$. The SM value is depicted as the dashed black line.}
    \label{fig:BR_Jpsi}
\end{figure}

Another interesting prediction of our models is the enhancement of the Higgs coupling to charm quarks. Although the detection of the Higgs decay to a charm pair probably has to wait for a linear collider, it might be possible to search for the Higgs decay to $J/\psi+\gamma$ at the High-Luminosity LHC.
In Figure~\ref{fig:BR_Jpsi} we present the BR$(h\to J/\psi+\gamma)$ for Type-A, B, I, and II. We note that the newly proposed types give the strongest enhancement above the SM value~\cite{Coyle:2019hvs}.

Finally, the most direct signature of any 2HDM is the discovery of the full Higgs spectrum at the LHC.
The main production mechanism of the heavy scalar, $H$, for $t_\beta\lesssim 10$ is gluon-fusion where the top-loop dominants the cross section.
However, for larger values of $t_\beta$ the contributions from the bottom-loop in Type-A and II or the charm-loop in Type-B have to be included.
In fact, the large enhancement for the bottom-Yukawa coupling arising in Type-A opens the possibility to consider the $b$-associated production of $H$. This has already been considered in the literature. For Type-B due to the enhancement of the charm-Yukawa coupling also $c$-associated production could become relevant. A detailed discussion of this aspect is beyond the scope of this paper.

\begin{figure}[t]
    \includegraphics[width=\linewidth]{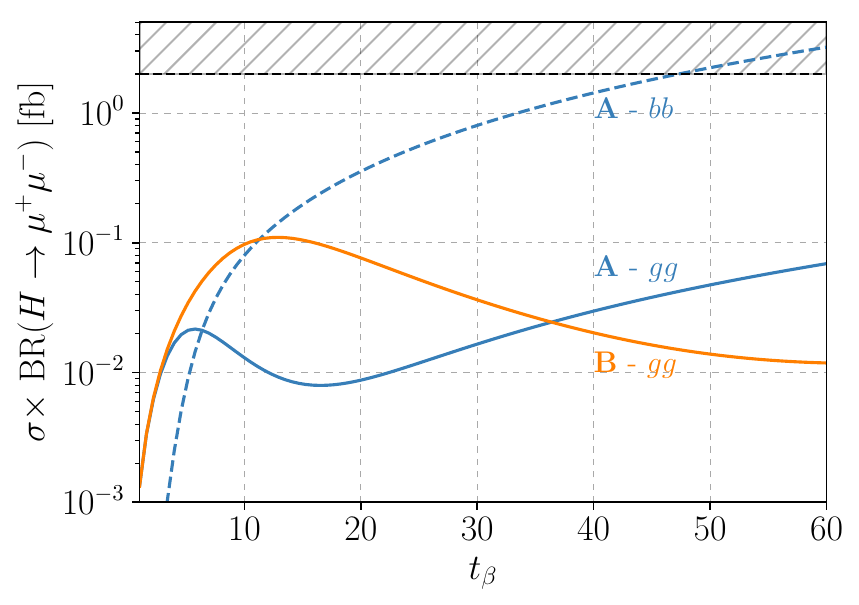}
    \includegraphics[width=\linewidth]{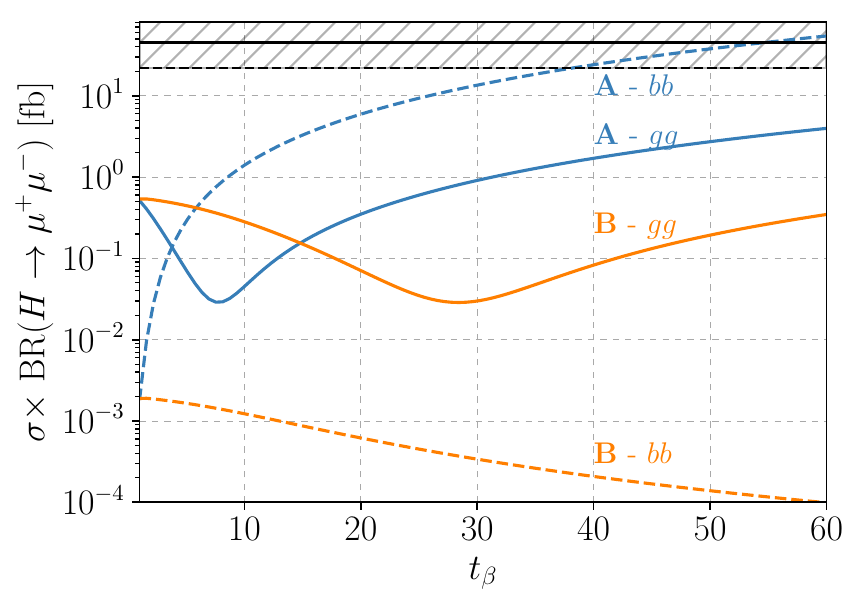}
    \caption{Values of $\sigma\times$BR$(H\to\mu^+\mu^-)$ with $c_{\beta-\alpha}=0$ for $M_H=500$\,GeV (top) and $M_H=250$\,GeV (bottom) in Type-A (blue) and B (orange) together with the corresponding limits from CMS (black)~\cite{Sirunyan:2019tkw}.
    The dashed (solid) lines indicate $b$-associated (gluon fusion) production. Contributions from $b$- and $c$-loops to the gluon fusion production are included.
    We note that Type-A is identical to Type-II in this channels.}
    \label{fig:H_mumu_exc}
\end{figure}

Searches for the heavy resonances decaying into muon pairs are of potential interest for our models. The recent ATLAS and CMS searches in~\cite{Aaboud:2019sgt, Sirunyan:2019tkw} distinguish between the gluon-fusion and $b$-associated production channels.
The results are present as upper limits on the production cross section, $\sigma$, times the BR$(H\to\mu^+\mu^-)$.
To obtain the production cross sections of $H$ we rescaled the NNLO results from Ref.~\cite{deFlorian:2016spz} to our parameter space.
For small to intermediate values of $t_\beta$ the suppression of $\xi^H_t$ is already effective but $\xi^H_{b(c)}$ is not strongly enhanced yet. For $m_H=500$\,GeV and $t_\beta > 10\, (50)$ the bottom (charm) contributions to gluon-fusion start to compensate the top coupling suppression in Type-A (B).
For Type-A the enhancement of $\xi^H_b$ is strong enough to exclude high values of $t_\beta$ in $b$-associated production.
These effects are shown in Figure~\ref{fig:H_mumu_exc}. There we adopted the slightly stronger upper bounds from CMS~\cite{Sirunyan:2019tkw} considering both production modes.

\begin{figure}[t]
    \includegraphics[width=\linewidth]{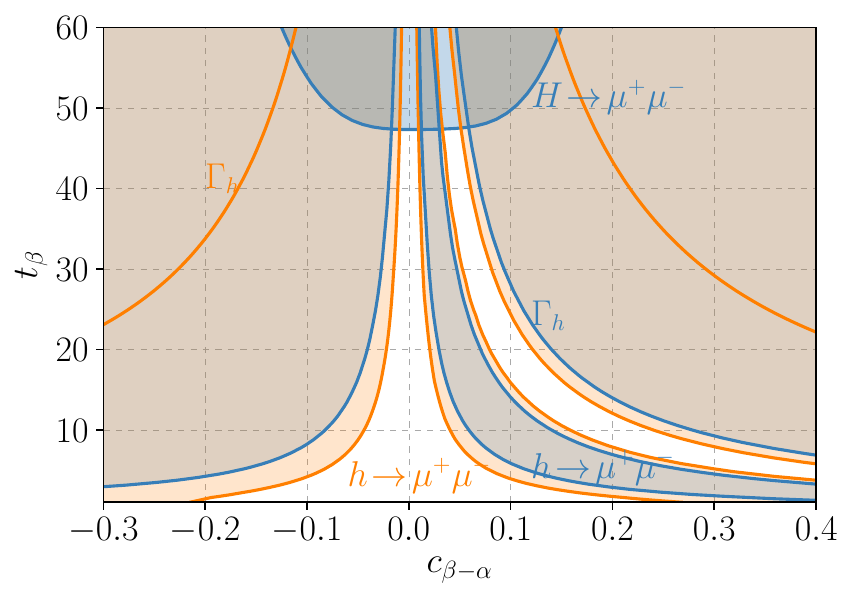}
    \caption{Summary of the discussed constraints on Type-A (blue) and B (blue) for $M_H=500$\,GeV.
    For Type-A we observe an interesting interplay of various measurements. For Type-B the dominant constraint arises from deviations of BR$(h\to\mu^+\mu^-)$.}
    \label{fig:summary}
\end{figure}

To summarize the constraints of special interest for Type-A and B we plot them together in Fig.\ref{fig:summary} in the ($c_{\beta-\alpha}, t_\beta$) plane. We find that even with this selection of channels large parts of the parameter space can be excluded.

\section{CONCLUSIONS} \label{sec:conc}
\noindent
Motivated by the mass hierarchy between the top quark and the other fermions, or between the third generation and the first two generation fermions, respectively, we investigated two new, interesting types of 2HDMs (called A and B) with a softly-broken $\mathbb{Z}_2$ symmetry.
Phenomenologically speaking, the new types are closely related to the well-known types of 2HDMs (I and II). This is a direct consequence of the great similarity between the models when only the heaviest fermions are considered. In return, that allowed us to comprehensively investigate their phenomenological viability and to make use of derived constraints to restrict the parameter space. Nevertheless, the proposed types feature deviations from the conventional ones in channels of potential reach of the HL-LHC enabling a distinction between them.

We studied the implications of assuming a flavor conserving ansatz, namely general flavor or singular alignment~\cite{Penuelas:2017ikk,Rodejohann:2019izm}.
By the virtue of it FCNCs are forbidden at tree-level. This allowed us to discuss new multi-scalar scenarios which address simultaneously the fermion mass hierarchy and the smallness of flavor violating processes mediated by neutral bosons.
The former aspect became possible through two VEV scales, $v_2^2 \gg v_1^2$.
Within the new types two naturalness criteria (Dirac and 't Hooft) might be realized in the scalar and/or Yukawa sector. In that sense we obtain more natural models.

Most of the low-energy and collider constraints derived for models of Type-I and II also apply to our two models.
As aforementioned, this is mainly a consequence of the strong similarity between our two types to the conventional ones. 
Therefore, we adopt constraints derived for them.
In particular, strong constraints originate from $b \to s$ flavor violating transitions which require the charged scalar mass to be above $600$\,GeV.
Given the phenomenological relevance of the AL we allowed at most small deviations from it.

However, specific signatures can be identified and used to distinguish our models from the conventional ones, namely
i) $h\to \mu^+ \mu^-$, ii) $H\to \mu^+ \mu^-$, and iii) $h \to J/\psi + \gamma$.
With the former decay it is possible to exclude large values of $t_\beta$ for Type-B, while for Type-A most of the range remains consistent with current data. On the other hand, for the latter decay we found that our two types give the strongest enhancement above the SM value compared to the conventional NFC scenarios.
Additionally, since the most direct signature of any 2HDM is the detection of the full scalar spectrum, we considered viable decay channels of the heavy CP-even neutral scalar. In particular, the decay into muon pairs can exclude large values of $t_\beta$ for Type-A even in the AL.

Overall, the architecture of the two newly proposed types offers new exciting possibilities to construct multi-Higgs models taking the observed hierarchies in the fermion mass spectrum into account and at the same time avoiding dangerous FCNCs in a natural manner. This is certainly an ambitious goal. Many issues should still be addressed to fully understand the pros and cons compared to the well studied conventional types.

\section*{ACKNOWLEDGMENTS}
\noindent
U.J.S.S. and J.L.D-C.~acknowledges support from CONACYT (M\'exico).	V.T.T.~acknowledges support by the IMPRS-PTFS. K.M.T.N.~acknowledges support from the research training group ``Particle Physics Beyond the Standard Model" (Graduiertenkolleg 1940). The authors thank the anonymous referee for the valuable insights that considerably improved the final version of this work.

To the memory of Prof. Alfonso Mondrag\'on, a man of heart and wisdom.

\bibliographystyle{apsrev4-1}
\bibliography{SA2HDM}

\end{document}